\documentclass[aa]{emulateapj}

\slugcomment{To appear in The Astrophysical Journal}

\shorttitle{Magnetic Topology of a Naked Sunspot: Is It really Nacked?}
\shortauthors{Sainz Dalda, A., Vargas Dom\'inguez, S. \& Tarbell, T. D.}

\begin{document}

\title{Magnetic topology of a naked sunspot: Is it really naked?}

\author{A. Sainz Dalda\altaffilmark{1}, S. Vargas Dom\'inguez\altaffilmark{2,3} 
       \& T. D. Tarbell\altaffilmark{4}}
\affil{$^1$ Stanford-Lockheed Institute for Space Research, Stanford University,
 HEPL, 466 Via Ortega, Stanford, CA 94305-4085, USA}
\affil{$^2$ Departamento de F\'isica, Universidad de Los Andes, A.A. 4976, 
Bogot\'a, Colombia}
\affil{$^3$ Mullard Space Science Laboratory, University College London, 
Holmbury  St Mary, Dorking, Surrey, RH5 6NT, UK}
\affil{$^4$ Lockheed-Martin Solar and Astrophysics Laboratory, 
Bld. 252, 3251 Hanover Street, Palo Alto, CA 94304, USA}

\clearpage

\begin{abstract}
The high spatial, temporal and spectral resolution achieved by Hinode 
instruments give 
much better understanding of the behavior of some elusive 
solar features, such as pores and naked sunspots. Their fast evolution and, in 
some
cases, their small sizes have made their study difficult. The moving magnetic 
features, despite being more dynamic structures, have been studied
 during the last 40 years. They have been always associated with sunspots, 
 especially with the penumbra. However, a recent observation of a naked sunspot 
(one with no penumbra)  
has shown MMF activity. The authors of this reported observation expressed their
 reservations about the explanation given to the bipolar MMF activity as an 
extension
of the penumbral filaments into the moat. How can this type of MMFs exist when a 
penumbra does not? In this paper, we study the {\it full} magnetic and 
(horizontal) 
velocity topology of the same naked sunspot, showing how the existence of a 
magnetic 
field topology similar to that observed in sunspots can explain these MMFs, even 
when 
the intensity map of the naked sunspot does not show a penumbra. 
\end{abstract}

\keywords{Sun: magnetic topology --- sunspots}

\clearpage

\section{Introduction}
The so-called {\it Moving Magnetic Features} (hereafter MMFs) were discovered  
by \cite{She69} and \cite{Vra71} and described in detail by \cite{Har73} and 
\cite{Vra74}. 
From the early observations, they were associated with the moat 
\citep{Vra74,Mey74,Bri88}: 
an annular region surrounding the sunspot where the MMFs runaway from the 
penumbra 
towards the network.  MMFs have been explained by  
an horizontal velocity field in the moat that drags the more horizontal, 
detached  
magnetic field lines from the bundle that forms the sunspot. 
Recently, MMFs have been observed coming from the mid-penumbra and entering 
 the moat region, which is dominated by large outflows \citep{Sai05, Rav06, 
Kub07}.
\cite{Sai05}  averaged a temporal sequence of SoHO/MDI high-resolution 
magnetograms  
that revealed a transverse component of the magnetic field beyond the penumbra  
outer edge as an extension of the most horizontal penumbral filaments in the 
moat.
High spatial resolution data have revealed a sea-serpent behavior of the more 
horizontal penumbral filaments as responsible for the bipolar magnetic 
structures
 in the mid-penumbra that become MMFs when they reach the moat \citep{Sai08}.
All these results establish a link between MMF activity and the horizontal 
magnetic  
field component in the penumbra with the Evershed flow, as was suggested early 
on by 
\cite{Vra74}.

One of the open questions about the moat is whether it exists or not around 
pores and naked  
sunspots\footnote{We consider a pore and a naked sunspot to be different  
solar features.  We understand a naked sunspot to be a solar feature that does 
not show a 
penumbra when it is observed, but that had or will have a penumbra during its 
life. A 
pore never develops a penumbra during its whole lifetime.}. 
 \cite{Var07, Var08} observed sunspots with different penumbral configurations 
and 
using {\it local correlation tracking} techniques (hereafter LCT) pointed out 
that 
the horizontal velocity flow is present in the part of the sunspot where 
penumbra exists. \cite{Var10} observed the horizontal flows around several 
pores. They did 
not find a moat around pores. On the contrary, they found an inflow region 
surrounding them. 
\cite{Bel08} analyzed the spectropolarimetric decay of a sunspot penumbra. They 
observed some finger-like structures remaining out of the naked sunspot after   
the penumbra disappeared. The presence or absence of these finger-like 
structures  
can be due to the evolutionary state of the (naked) sunspot. In our study we do 
not see
 these signatures, but the strength and the inclination of their magnetic field 
values in and out of the naked sunspot are very similar (see Figure 3 in \citealt{
Bel08}). Here, we present a similar one-instance study of the magnetic topology 
of  
the naked sunspot observed by \cite{Zuc09} (hereafter ZETAL09). 

ZETAL09 found MMF activity around a naked sunspot, asking for a
 revision of the previous results that related this activity to the horizontal
 magnetic field. Their main conclusion is: 
`{\it The presence of bipolar MMFs in a naked spot indicates that current 
interpretation  
of bipolar MMFs, as extensions of the penumbral filaments beyond the sunspot 
outer 
boundaries, should be revised, to take into account this observational 
evidence}'.

In this letter, we use spectropolarimetric measurements to present a new 
perspective  
on these data. As results, the MMFs are again related to the horizontal   
magnetic field component of the (naked) sunspot, usually associated with the  
penumbra, but not necessarily. We also give a new interpretation of the LCT 
results that shows
an horizontal velocity flow distribution in agreement with previous results.

\section{Data as viewed from the magnetic field}
The AR NOAA 10977 was observed in the  Fe {\small I} 6301 and 6302 \AA\ lines 
from 
16:25 to 16:47 UT on 5 December 2007 with the spectropolarimetric instrument  
SOT/SP \citep{iTsu08} on the Hinode satellite \citep{Kos07}. 
This is the closest available in time to the data studied by ZETAL09 (14:06 to 
15:48 UT). The  
spatial and spectral sampling were 0\arcsec.32 and 22 $m$\AA\ respectively. At
 the observation time, the naked sunspot was located at 
heliographic coordinates $(-12^{\circ}, -5^{\circ})$. To obtain the most 
accurate  
values of the physical parameters we have applied the SIR inversion code 
\citep{Bas92} to the Stokes profiles. 
The calibrated profiles are easily obtained thanks to the data reduction tools 
mainly developed  
by B. Lites and available in the SOT {\it SolarSoft} package. 
Several inversions with different initializations were done. Here, we present 
one that represents a 
trade-off between the accuracy in inferred values (i.e., with smallest 
errors) and the degrees of freedom allowed in the inversion. Therefore, we have 
chosen a  
combination of degrees of freedom that shows the best fit between the observed 
and the inverted 
profiles and is compatible with a reliable  atmosphere model. We have used a 
simple model 
with a magnetic component occupying the whole pixel (i.e., filling factor is 1.0) 
and a fixed 
stray light contribution of 15\%. We have verified that an inversion with stray 
light as 
a free  parameter does not introduce significant improvement in the output 
model; 
therefore we selected a high mean value in the studied region,
and kept it fixed in the inversion presented here. 

Figure~\ref{fig_maps} shows maps of the naked sunspot belonging to the AR NOAA
10977 after the calibration of the Stokes parameters (first row) and their inversion
 (second and third row). In the top row, we present the
Stokes Intensity (with respect to the continuum intensity) map, the Mean
Circular Polarization Degree map (hereafter MCPD map) and the Mean Linear Polarization
Degree map (hereafter MLPD map). The former two maps were respectively
calculated as $MCPD = \int_{\lambda_{0}}^{\lambda_{1}}{\frac{|V(\lambda)|d\lambda 
}{I(\lambda)}}$ and $MLPD = \int_{\lambda_{0}}^{\lambda_{1}}{\frac{\sqrt{Q^2(\lambda) + 
U^2(\lambda) }}{I(\lambda)}d\lambda}$, being $\lambda_{0}$ and $\lambda_{1}$
6302.27$\pm0.02$ \AA\ and  6302.70$\pm0.02$ \AA\ respectively, i.e., a spectral range including the
line \ion{Fe}{1} 6302 \AA. The MCPD is a good proxy for the unsigned vertical
component of the vector magnetic field, while the MLPD is good for the transverse
component of the vector magnetic field.  In these maps we have overlaid the contours
corresponding to the displayed magnitude. Thus, the intensity contour at level $I/I_{c} = 0.8$
delimits the naked sunspot from the granulation. In this paper, we shall refer
to this region as \begin{it}photometric naked sunspot,\end{it} in the sense that
its nature is uniquely described by the intensity magnitude. The contour at the MCPD
map encloses the area where $ MCPD > 3.5\% $. Similarly, the contours at MLPD map
delimit the area where $MLPD > 1\%$. In these maps, the intensity contour is displayed
(black) as reference. Notice that in the MLPD maps there is a contour inside the
region delimited by the intensity contour, which belongs to the MLPD contour.
These three maps simply retrieved from the Stokes profiles offer us valuable
information at a glance. The most obvious is the existence of an extended
magnetic field beyond the photometric naked sunspot edge.
Both the longitudinal and the transverse component of the magnetic field are
present inside the photometric naked sunspot and beyond its intensity contour. However,
although valuable, these maps are a rough approximation to the vector magnetic field: they can not
tell us much about either the strength of the magnetic field or its {\it true} topology.

The second row of Figure~\ref{fig_maps} shows the strength (left), the vertical 
(center, 
$B_{vert}$) and the horizontal (right, $B_{hor}$) components of the magnetic  
field vector\footnote{Whereas the terms longitudinal and transverse are used for 
the projection of the magnetic field vector on the observation reference frame, 
the terms vertical and horizontal are used for the projection on the local reference
 frame (hereafter LRF).}. In the strength map, three contours have been overlapped, 
 from the inner to the outer: the intensity contour (black), the contour for 
values  where $B_{vert} > 0.25$ kG (white) and the contour for values where 
$B_{hor} 
> 0.15$ kG (white). The last two contours are respectively shown in the $B_{vert
}$ and $B_{hor}$ maps. $B_{vert}$ values drop from roughly $1.10\pm0.07$ kG at 
the 
edge of the photometric naked sunspot to $0.25\pm0.03$ kG at the position of the 
$B_ {vert}$ contour, which is roughly located $1\arcsec$ outside of the 
intensity 
contour of the naked sunspot. $B_{hor}$ values go from $0.80\pm0.05$ kG at the 
outer part of the photometric naked sunspot to $0.15\pm0.03$ kG at $1.5\arcsec$ 
outside of the intensity contour where the $B_{hor}$ 
contour is located on average. 

Finally, the third row shows the inclination in the LRF (left), velocity along 
the 
line of sight (hereafter LOS, center) and temperature (right). 
In this row, all maps have been overlaid with the three contours (intensity, 
$B_{vert}$ and $B_{hor}$). 
At the center of the naked sunspot the inclination is $180^{\circ}$, i.e., the 
magnetic field is directed inward to the solar surface. 
The inclination of the magnetic field in the LRF roughly drops from 
$140\pm4^{\circ}$ ($50\pm4^{\circ}$ with respect to the local horizontal) to 
$110\pm4^{\circ}$ ($20\pm4^{\circ}$) out the photometric naked sunspot. The LOS 
velocity is upward or close to zero everywhere both the photometric naked 
sunspot and
its surroundings: there is not any trace of Evershed flow, even though the 
position of 
the naked sunspot is far off the center of the solar disc. The temperature map 
shows 
a hot ring just between the intensity contour and $B_{vert}$ contour. The 
temperature 
of this ring is roughly $5.50\pm0.08$ kK. It is hotter than the temperature of 
the photometric naked sunspot ($4.00\pm0.03-5.00\pm0.05$ kK) but cooler than the 
most prominent granules in the studied area. We can see one granule located at 
map 
coordinates (2,10) with a temperature of $5.50\pm0.03$ kK. The other granule is 
located at map coordinates (12,3), and it shows a temperature similar to the 
former  
one but in this case the error is $\pm0.10$ kK. On average, the temperature of 
this 
ring is slightly hotter than the granulation, but it is still 
cooler than the hotter granules shown in the map. 

The values of both components of the magnetic field are stronger than the noise 
level, and the errors of the components and inclination of the vector magnetic 
field 
presented in this paper are only slightly higher than the maximum errors 
obtained by \cite{Gos10}. 
They calculated an error value for vector magnetic field of a sunspot observed 
by 
Hinode-SOT/SP using both a Monte-Carlo approach and a Milne-Eddington inversion. 
The maximum values that they obtained were: $\pm 50$ G for the field strength 
and 
$\pm3^{\circ}$ for the inclination. Therefore, our results look as reliable and 
consistent as other observations and inversions (see also result and errors in 
\citealt{Bel08}). 
To summarize, the topology of the (full) vector magnetic field of the observed 
naked sunspot has been revealed. We refer to to this new vision of the naked 
sunspot as the {\it magnetic naked sunspot}.  
  
\section{Data as viewed from the velocity field}
In order to analyze the proper motions around the (naked) sunspot, we have 
selected  
the same SOT/BFI data used by ZETAL09 for applying LCT to compute the flow map  
of horizontal velocities. A total of 51 G-band images from 14:06 to 15:48 UT 
($\sim$2 min cadence) were processed with standard \emph{SolarSoft} routines and  
co-aligned at a sub-pixel level by cross-correlation over the entire FOV between subsequent
pairs of images. A subsonic filter (velocity threshold  
of 4 km s$^{-1}$) was used to eliminate p-modes resulting in a final time series  
of 46 images after apodization.

The LCT procedure was then applied by using a correlation tracking window of 
FWHM 1\farcs0. Two cases (Cases 1 and 2) of maps of 
horizontal velocities were generated. In the computed map for Case 1 (same as 
presented by ZETAL09) 
the velocity vectors have a predominant trend to the right, 
and the velocities closer to the right edge of the FOV are generally larger  
in magnitude. Figure~\ref{flowmaps} (top left panel) shows a clipped region of
 the FOV in the vicinity of the (naked) sunspot. In Case 2, the time series  
were first aligned with respect to a window enclosing the (naked) sunspot. By 
doing this, we neglect the proper (inherent) motion of the sunspot through the 
surrounding granulation, thus focusing on the plasma motions around the 
\emph{anchored} sunspot. We detected shifts with respect to Case 1 of up to 26 
and 10 pixels in  
the x and y directions respectively. Figure~\ref{flowmaps} (top right panel) 
shows the resulting map for Case 2 for comparison with Case 1 (left). We follow 
the same procedure as used by Vargas Dom\'inguez (2010) to establish the 
direction of velocity vectors around the (naked) sunspot. Figure~\ref{flowmaps} 
(bottom panels) shows binary maps displaying the distinction between inward 
(white) and outward (black) 
radial components of the velocity vectors for Case 1 (left) and Case 2 (right) 
respectively. Our results agree with previous results for pore-like structures 
(Vargas Dom\'inguez, 2010) showing that motions towards the pore are dominant  
in the closest vicinity. For Case 2 this behavior is even more evident and 
symmetric around the (naked) sunspot, showing the differential proper motions of 
plasma around the \emph{anchored} spot.  Motions at the periphery of
the structure are significantly influenced by external plasma flows caused by 
exploding events as observed in previous works, for example, \citep{Sob99}.

\section{Discussion, Criticism  and Conclusions}
We remark that a moat and MMF activity could be considered as observed magnetic 
features, while the umbra and penumbra are observed intensity (or photometric)  
features. However we should not forget that this classification is based on the 
primary technique used for their identification, and several additional aspects 
of their nature should also be considered.  

Figure~\ref{fig_maps} clearly shows a prominent horizontal
magnetic component around the photometric naked sunspot.  This horizontal 
component is co-spatial
with the vertical component of the magnetic field in the outermost part of the 
photometric naked sunspot, and it persists as an horizontal magnetic structure 
beyond. 
This magnetic field configuration is very similar to that of a 
standard sunspot, at least from the magnetic point of view. That is, if we
understand a sunspot as an intensity structure, the sunspot studied by ZETAL09 
can be classified as a so-called {\it naked sunspot}. On the other hand, the 
configuration 
of the magnetic field resembles that of a sunspot with extended magnetic
field beyond the penumbra \citep{Sai05} and of a naked 
sunspot during the decay of a sunspot \citep{Bel08}. 
\cite{Sai08} using Hinode/NFI magnetograms sketched a possible scenario where 
the sea-serpent behavior of the more horizontal penumbral filaments explain the 
bipolar magnetic structures in the mid-penumbra that eventually become MMFs when
 they reach the moat.  Both observations related the more horizontal magnetic 
field component of the sunspot with the MMF activity, in agreement with several 
observational and theoretical proposals \citep{Har73, Schl02}. In the data 
presented here, we observe a very similar magnetic field configuration, although 
the sunspot does not have a photometric penumbra.  It has a very similar 
magnetic structure surrounding the naked sunspot, although without a filamentary 
distribution at the spatial resolution in these observations.  
The main conclusion presented by ZETAL09 was based on the observational evidence  
that the selected sunspot is a naked sunspot. That is true but only part of the 
story: only taking into account the intensity point of 
view, applying LCT without correcting for the inherent proper motion of the 
(naked) sunspot and focusing only on the proper 
motions of the surrounding granulation. In our analysis we have demonstrated 
that the apparent outflows from the naked spot are not actually moat flows (as 
suggested by ZETAL09), but rather the contribution from outward flows 
originating in the 
regular mesh of divergence centers around the pore, in agreement with previous 
results 
 \citep[e.g.,][]{Var10}. 

They did not take into account the whole, true magnetic configuration of the 
sunspot. 
They related an intensity feature (sunspot without penumbra) with a magnetic 
structure 
(MMF) only looking at the intensity and a magnetogram (i.e., longitudinal 
apparent 
magnetic flux map), neglecting to verify what the true magnetic 
topology of the sunspot was. 
Although ZETAL09 reported Ð- probably for the first time -- MMF activity  
around a naked sunspot, their suggested revision of the relationship between MMF 
activity  
and the extension of the more horizontal penumbral filaments on the moat can
be questioned if one considers the true (vector) magnetic field of the naked 
sunspot.
 Does it mean that {\it all} MMF activity can be explained {\it uniquely} by the
 extension of the horizontal penumbral filaments in the moat or even by an 
horizontal 
component of the magnetic field around the naked sunspot?  It does not, but the 
absence  of a (photometric) penumbra does not imply a lack of an horizontal 
field 
around naked sunspot, and therefore {\it some} MMF activity can still be 
explained 
by the current interpretation. So, the MMF of type I and III (see classification 
in 
\citealt{Shi00} and Figure 3 of \citealt{Tho02}) are compatible with the observed 
horizontal component around sunspots and (photometric) naked sunspots. 
We have shown how a (photometric) naked sunspot has associated a vertical and 
horizontal magnetic component slightly outside of its intensity edge and an 
horizontal  
magnetic component extending much further beyond its intensity edge. In this 
sense, the 
(magnetic) naked sunspot is a rather chaste sunspot. 

\section{Acknowledgments}
ASD thanks the International Space Science Institute  
and Irina Kitiashvili for the invitation to participate in the meeting 
`Filamentary Structure and Dynamics of Solar Magnetic Fields', 
15-19 November 2010, ISSI, Bern, where many aspects of this paper
were discussed with other colleagues. We also thank V. Mart\'inez Pillet and
B. Ruiz Cobo for their helpful comments.  
Hinode is a Japanese mission developed by ISAS/JAXA, with NAOJ as
domestic partner and NASA and STFC (UK) as international partners. It is operated 
in cooperation with ESA and NSC (Norway). The Hinode project at Stanford and Lockheed 
is supported by NASA contract NNM07AA01C (MSFC).

\clearpage

{\begin{figure} 
\includegraphics[width=\textwidth]{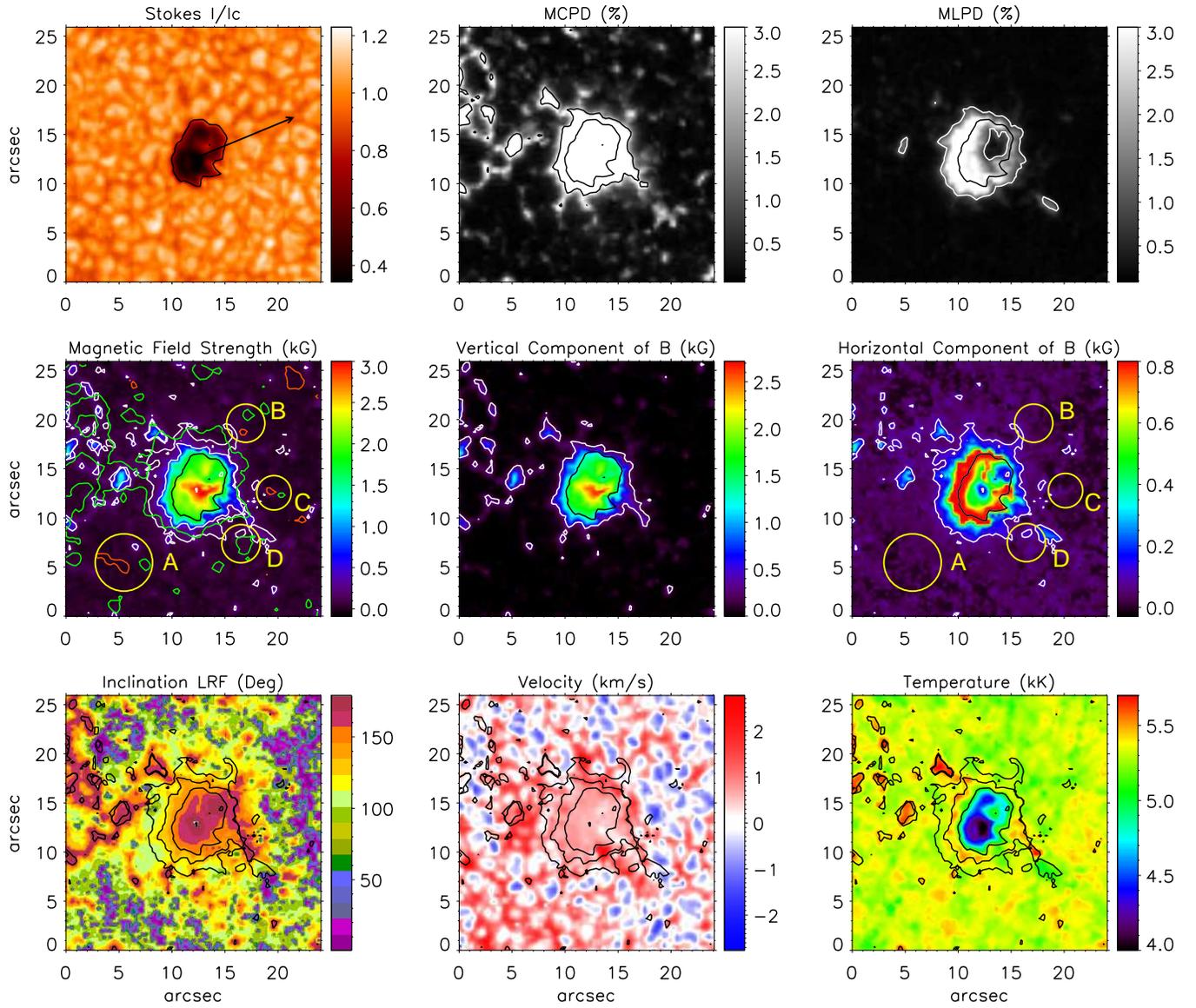}
\caption{\emph{Top panels}: intensity, mean circular polarization degree and
mean linear polarization degree maps calculated directly from the Stokes
parameters. \emph{Bottom panels}: some  physical values of the atmosphere
 obtained after the inversions done by SIR code. The second inner
contour delimits the area mainly related with the vertical component of the
magnetic field vector. The inner- and outermost contours delimit
the horizontal component of the magnetic field vector.}
\label{fig_maps}
\end{figure}}

\begin{figure}
\hspace{-5.5cm}\includegraphics[angle=90,width=1.7\linewidth]{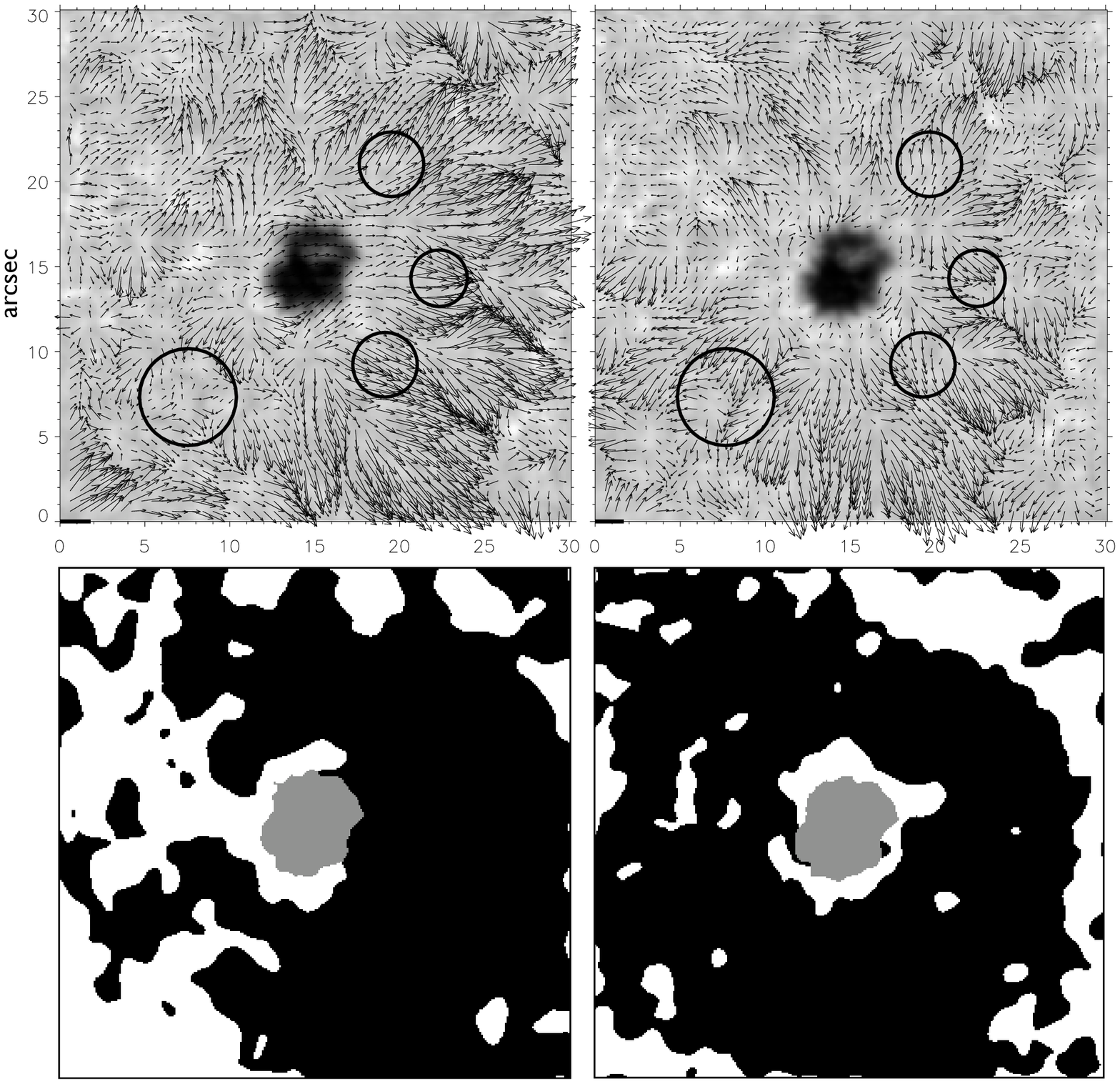}
\vspace{-1cm}
\caption{Distribution of horizontal proper motions around the naked (sunspot). 
\emph{Top panels}: Maps of horizontal velocities for the \emph{Hinode} (G-band) 
time series (92 min average) for Case 1 (left) and Case 2 (right). The background
 images represent the average image of the series. The length of the black bar 
at coordinates (0,0) corresponds to 2 km s$^{-1}$. \emph{Bottom panels}: Binary 
maps of inward (white) and outward (black) radial velocity components for Case 1 
(left) and Case 2 (right). See text for details.}
\label{flowmaps}
\end{figure}

\end{document}